\documentclass[conference]{IEEEtran}
\IEEEoverridecommandlockouts
\usepackage{cite}
\usepackage{booktabs} 
\usepackage{amsmath,amssymb,amsfonts}
\usepackage{algorithmic}

\usepackage{multirow}
\usepackage{amsmath}
\usepackage{array}
\usepackage[normalem]{ulem}
\usepackage{bbding}
\usepackage{url}

\usepackage{graphicx}
\usepackage{textcomp}
\usepackage{xcolor}
\def\BibTeX{{\rm B\kern-.05em{\sc i\kern-.025em b}\kern-.08em
    T\kern-.1667em\lower.7ex\hbox{E}\kern-.125emX}}

\IEEEoverridecommandlockouts\IEEEpubid{\makebox[\columnwidth]{979-8-3315-4940-4/25/\$31.00~\copyright~2025 IEEE \hfill} \hspace{\columnsep}\makebox[\columnwidth]{ }}

\begin{document}


\title{TraGe: A Generic Packet Representation for Traffic Classification Based on Header-Payload Differences\thanks{\IEEEauthorrefmark{6} Both authors contributed equally to this research.}\thanks{\IEEEauthorrefmark{1} Corresponding author.}}

\author{
  \IEEEauthorblockN{Chungang Lin\IEEEauthorrefmark{2}\IEEEauthorrefmark{3}\IEEEauthorrefmark{6}, Yilong Jiang\IEEEauthorrefmark{2}\IEEEauthorrefmark{3}\IEEEauthorrefmark{6}, Weiyao Zhang\IEEEauthorrefmark{2}, Xuying Meng\IEEEauthorrefmark{2}\IEEEauthorrefmark{4}\IEEEauthorrefmark{1}, Tianyu Zuo\IEEEauthorrefmark{3}, Yujun Zhang\IEEEauthorrefmark{2}\IEEEauthorrefmark{3}\IEEEauthorrefmark{5}\IEEEauthorrefmark{1}}
  \IEEEauthorblockA{\IEEEauthorrefmark{2}Institute of Computing Technology, Chinese Academy of Sciences, China.}
  \IEEEauthorblockA{\IEEEauthorrefmark{3}University of Chinese Academy of Sciences, China.}
    \IEEEauthorblockA{\IEEEauthorrefmark{4}Purple Mountain Laboratories, China.}
    \IEEEauthorblockA{\IEEEauthorrefmark{5}Nanjing Institute of InforSuperBahn, China.}

  \IEEEauthorblockA{\{linchungang22s, jiangyilong23s, zhangweiyao17z, mengxuying, zuotianyu22s, nrcyujun\}@ict.ac.cn}
}

\maketitle

\begin{abstract}
Traffic classification has a significant impact on maintaining the Quality of Service (QoS) of the network.
Since traditional methods heavily rely on feature extraction and large-scale labeled data, some recent pre-trained models manage to reduce the dependency by utilizing different pre-training tasks to train generic representations for network packets.
However, existing pre-trained models typically adopt pre-training tasks developed for image or text data, which are not tailored to traffic data.
As a result, the obtained traffic representations fail to fully reflect the information contained in the traffic, and may even disrupt the protocol information.
To address this, we propose TraGe, a novel generic packet representation model for traffic classification.
Based on the differences between the header and payload—the two fundamental components of a network packet-we perform differentiated pre-training according to the byte sequence variations (continuous in the header vs. discontinuous in the payload). A dynamic masking strategy is further introduced to prevent overfitting to fixed byte positions.
Once the generic packet representation is obtained, TraGe can be fine-tuned for diverse traffic classification tasks using limited labeled data.
Experimental results demonstrate that TraGe significantly outperforms state-of-the-art methods on two traffic classification tasks, with up to a 6.97\% performance improvement.
Moreover, TraGe exhibits superior robustness under parameter fluctuations
and variations in sampling configurations.
\end{abstract}


\section{Introduction}

Traffic classification aims to organize network traffic into different categories (e.g., application and service), which is a fundamental and vital technology in network management and Quality of Service (QoS) \cite{IWQOS1,IWQOS2,IWQOS3,ICC24,etbert,YaTC}.
{For instance, in an enterprise network, traffic may include email, voice calls, and file transfers. By effectively classifying these traffic categories, network administrators can implement QoS policies to ensure critical traffic like voice calls receive sufficient bandwidth, low latency, and prioritized network resources \cite{QoS1,CN25}.}

Traditional statistical feature-based methods \cite{FlowPrint,AppScanner,DTree,XGBoost}, such as FlowPrint \cite{FlowPrint} and AppScanner \cite{AppScanner}, rely on manual traffic feature extraction, where the extracted features are then fed into machine learning models.
In contrast, deep learning-based methods \cite{Fs-Net,EBSNN,TFEGNN} automate the learning of traffic features through deep learning models, reducing the need for manual feature extraction.
However, deep learning models depend heavily on large-scale labeled data.
To address this, some recent pre-trained models \cite{PERT,etbert,NetGPT,YaTC} leverage large-scale unlabeled raw byte sequences from packets, rather than extracted statistical features, to learn generic representations, requiring only small-scale labeled data for fine-tuning and thereby reducing the reliance on both feature extraction and large-scale labels.

However, existing pre-trained models mainly apply pre-training tasks originating from other domains, such as computer vision (CV) and natural language processing (NLP). These pre-trained tasks are tailored to domain-specific data formats—image data in CV and text data in NLP—which differ significantly from the traffic data in the network domain, making them less effective in learning a generic representation for network traffic.
A network packet comprises two fundamentally components: the header and the payload.
In contrast to image or text data, which are typically modeled as uniform data, the header and payload exhibit significant structural differences.
Specifically, unlike the payload, the header contains structured protocol fields with continuous byte sequences that are essential for recognizing traffic patterns. For instance, the TCP sequence number can reveal program behavior and support anomaly detection \cite{HBC}. 
Yet, existing pre-trained models overlook the header byte continuity by either classifying traffic based solely on the payload (e.g., PERT \cite{PERT}, ET-BERT \cite{etbert}) or by treating the header and payload uniformly, thus disrupting this continuity (e.g., NetGPT \cite{NetGPT}, YaTC \cite{YaTC}).

To address the above limitations, in this paper, we propose TraGe, a novel pre-trained model for traffic classification with a generic packet representation.
This representation is derived by fully considering the header-payload differences of network packets.
Specifically, TraGe performs differentiated pre-training based on the byte sequence variations (continuous vs. discontinuous) between the header and payload, with a dynamic masking strategy applied to avoid overfitting to fixed byte positions.
Together, these components enable TraGe to obtain a generic packet representation, which can then be fine-tuned for various traffic classification tasks with limited labeled data.
Extensive experiments conducted on publicly available datasets demonstrate the effectiveness and the robustness of TraGe. 
The comparison results show that TraGe outperforms state-of-the-art pre-trained models, achieving up to a 6.97\% improvement in classification performance. In addition, analyses of parameter sensitivity and sampling variability demonstrate TraGe's robustness to both parameter fluctuations and variations in sampling configurations.

The main contributions of this paper are as follows:

\begin{itemize}
    \item We propose TraGe, a novel pre-trained model with a generic packet representation for traffic classification based on header-payload differences.
    \item We design two pre-training tasks (i.e., MLM-FM and MLM-RM) for the header and payload of network packets to obtain generic packet representations.
    \item We conduct extensive experiments on two traffic classification tasks to evaluate the effectiveness and the robustness of TraGe.
\end{itemize}

\section{Related Work}

According to the utilized features, model structure, and pre-training technology, related works can be divided into methods based on statistical features, deep learning, and pre-training technology.

\subsection{Methods based on Statistics Features}
Statistical feature-based methods rely on manually extracting traffic features for model training. For example, FlowPrint \cite{FlowPic} uses packet inter-arrival times and applies clustering with cross-correlation for classification. AppScanner \cite{AppScanner} trains a random forest classifier using packet size statistics, while DTree \cite{DTree} incorporates size and timing features in a decision tree model. Although these methods demonstrate reasonable effectiveness, their reliance on handcrafted features restricts the model’s ability to learn comprehensive network patterns.

\subsection{Methods based on Deep Learning}
Deep learning-based methods automate feature extraction by directly processing raw traffic data through neural networks, eliminating the need for handcrafted features. For example, FS-Net \cite{Fs-Net} uses recurrent neural networks (RNNs) to learn representations from raw packet length sequences. EBSNN \cite{EBSNN} also leverages RNNs but instead operates on raw byte sequences, with two variants: EBSNN-L (based on LSTM) and EBSNN-G (based on GRU). TFE-GNN \cite{TFEGNN} further explores this direction by modeling byte sequences as graphs and applying graph neural networks (GNNs) for classification. Despite their effectiveness, these methods typically demand large amounts of labeled data, which is often impractical in real-world deployment scenarios \cite{NetAugment,PERT,etbert}.

\subsection{Methods based on Pre-training Technology}
Pre-trained models leverage self-supervised learning on large-scale unlabeled traffic to learn generalizable representations, which are then fine-tuned using a smaller set of task-specific labeled data. This strategy reduces reliance on extensive annotations by enabling effective classification with limited labeled samples. For example, ET-BERT \cite{etbert} treats payload byte sequences as word-like inputs and applies two NLP-inspired pre-training tasks: Masked Language Modeling (MLM) and Next Sentence Prediction (NSP). Likewise, YaTC \cite{YaTC} models packet byte sequences—including headers and payloads—as images and uses Masked Image Modeling (MIM), a task from the computer vision domain, to capture inter-traffic relationships. However, current pre-trained models predominantly adopt pre-training tasks from CV and NLP, which are designed for structured image or text data. These data types differ significantly from network traffic, limiting the effectiveness of such tasks in capturing traffic-specific representations.


\section{Methodology}

We propose TraGe to fully exploit the traffic characteristics for learning general packet representations, thereby improving model performance in different network traffic classification tasks.
{
Specifically, TraGe leverages the explicit byte distribution differences between the header and the payload to perform header-payload differentiated pre-training (Section \ref{sec:Header-Payload Differentiated Pre-training}). This approach enables the model to obtain a generic packet representation, which facilitates effective model fine-tuning (Section \ref{sec:Model Fine-tuning}) across diverse traffic classification tasks. The overall framework of TraGe is shown in Fig. \ref{fig:The overview of TraGe}.}

\subsection{Header-Payload Differentiated Pre-training}
\label{sec:Header-Payload Differentiated Pre-training}

\begin{figure}[t!]
\centerline{\includegraphics[width= 0.98 \linewidth]{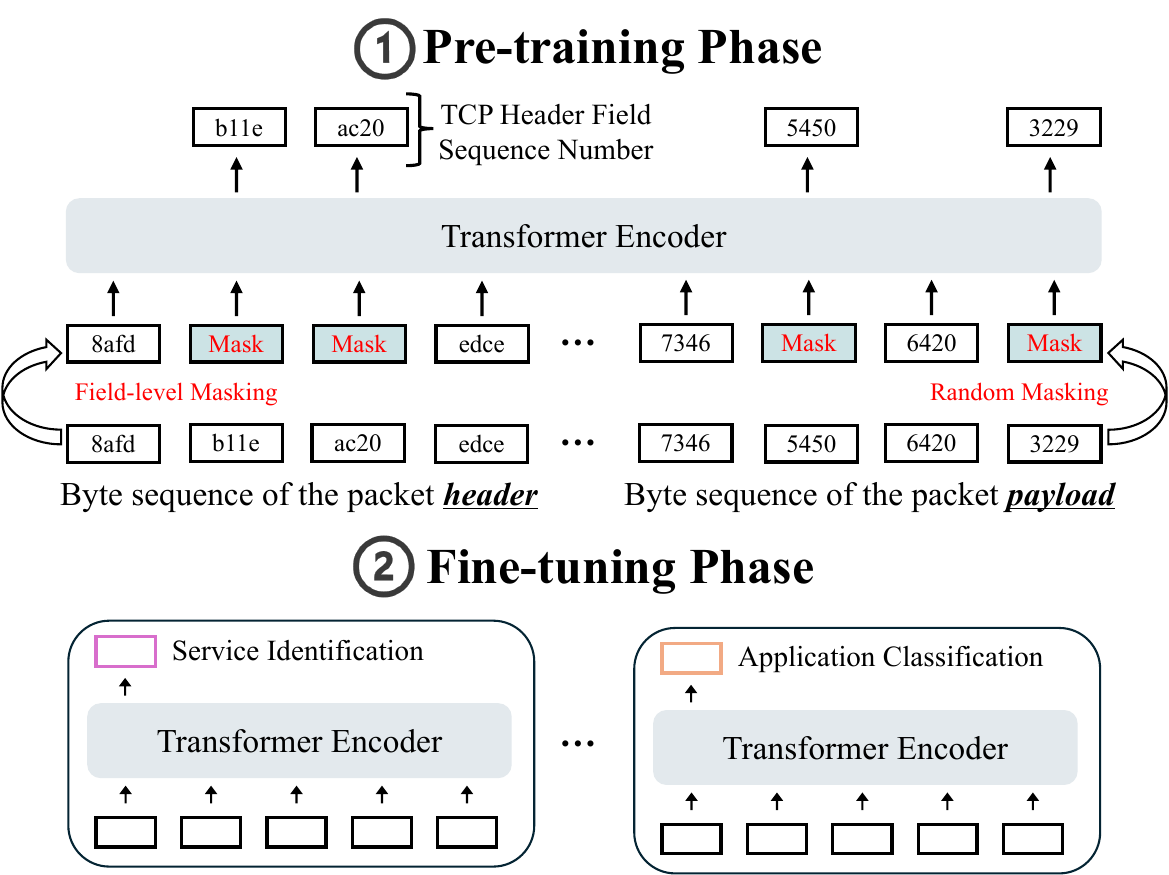}}
\vspace{-7pt}
\caption{The framework of TraGe.}
\vspace{-10pt}
\label{fig:The overview of TraGe}
\end{figure}

The header and payload of a packet both consist of byte sequences, but the same byte sequence can represent different meanings.
For instance, as shown in Fig. \ref{fig:The overview of TraGe}, the byte sequence “b11eac20”\footnote{We represent byte sequences in lowercase hexadecimal format.}  denotes a sequence number in the header, indicating the position of the packet within a TCP flow.
In contrast, in the payload, “b11eac20” could represent a specific data value or instruction, depending on the context of the application and protocol.
If identical byte sequences in the header and payload are processed uniformly without considering their distinct meanings, the model may struggle to correctly extract relevant information, which can hinder its learning performance.


Given the differences between the header and payload, we propose header-payload differentiated pre-training based on the byte distribution (continuous vs. discontinuous), as shown in Fig. \ref{fig:The overview of TraGe}.
{
This approach introduces two pre-training tasks—MLM-FM for the header and MLM-RM for the payload—which together form the foundation of the final generic packet representation. To further enhance pre-training effectiveness, we incorporate a Dynamic Masking (DM) strategy to mitigate model overfitting.}

\paragraph{Pre-training Task for Header} The network packet header comprises protocol header fields, with each field consisting of \textit{continuous} byte sequences.
Existing pre-training models primarily employ Masked Language Modeling (MLM) pre-training task \cite{etbert,PERT}, where random byte sequences are masked during model pre-training.
The model then learns to predict these masked bytes, thereby forming a generic packet representation.
However, the random masking approach disrupts the continuity of the header fields. 
For example, a byte sequence like “b11eac20” may represent specific network information, such as a sequence number.
Random masking could mask only part of the sequence (e.g., just the byte “b1”), preventing the model from fully capturing the intended network context.
As a result, the model’s ability to learn a generic packet representation is compromised.

\begin{figure}[h]
\vspace{-2pt}
\centerline{\includegraphics[width=1 \linewidth]{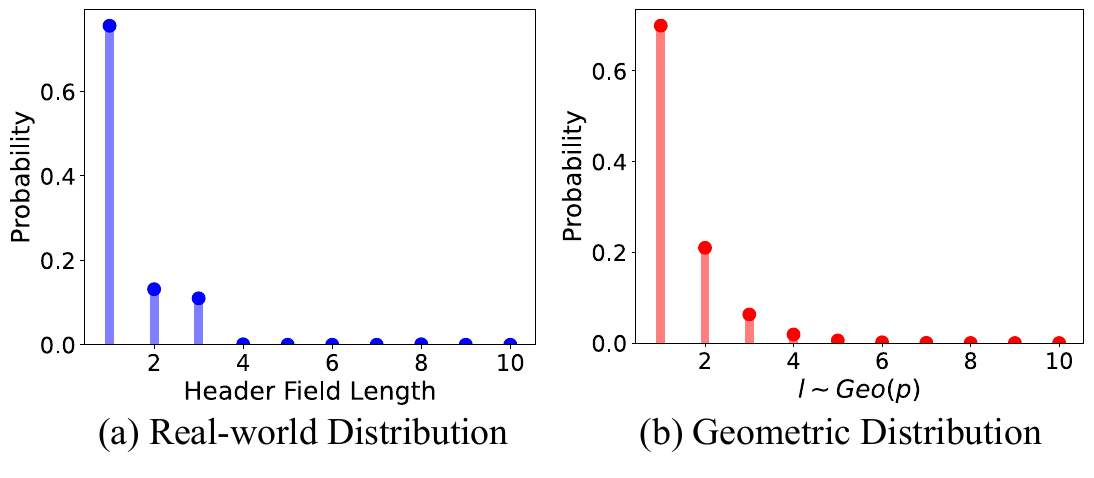}}
\vspace{-8pt}
\caption{A case study on header field length distribution on the ISCX-VPN dataset and geometric distribution.}
\label{fig:case}
\end{figure}

Therefore, we introduce field-level masking to continuously mask byte sequences, thereby improving the model's ability to learn generic packet representations.
A feasible approach to implementing field-level masking is to parse each packet’s header in advance to identify the location of each protocol header field, followed by masking the relevant fields. 
However, parsing packets can be computationally expensive, and not all protocol headers are easily parsed.
To address this, we propose an alternative approach using a geometric distribution.
Specifically, we sample the length of the masked byte sequence based on a geometric distribution, drawing inspiration from the observed similarity between the length distribution of header fields and a geometric distribution.
As shown in Fig. \ref{fig:case}, the distribution of header protocol field lengths in the ISCX-VPN \cite{ISCX-VPN} dataset closely aligns with the geometric distribution $\ell \sim \operatorname{Geo} (p)$.
This similarity supports the feasibility of using a geometric distribution for field-level masking.\footnote{We evaluate the impact of the geometric distribution parameters on the field-level masking in Section IV-D. The experimental results show insignificant performance changes across these parameters.} 

During the pre-training of the transformer encoder for the packet header, we first sample a length  $l$  using a geometric distribution $\ell \sim \operatorname{Geo} (p)$.
Next, we continuously select  $l$  tokens from the input sequence, with each token consisting of two consecutive bytes, following the approach in ET-BERT \cite{etbert}.
As shown in Fig. \ref{fig:The overview of TraGe}, for packet \#$k$, these tokens are masked with the special token [MASK],
and the transformer encoder is trained by predicting the masked tokens based on the context.
The loss function is based on negative log-likelihood, formally defined as:
\begin{align}
L_{\textit{MLM-FM}} = -\sum_{i = 1}^{l} \log \left(P\left(\textit{MASK}_{i} = { token}_{i} \mid {\theta}_{H} \right)\right)
\end{align}
where ${\theta}_{H}$ represents the model parameters of the encoder. We pre-train the transformer encoder through Masked Language Modeling (MLM) based on Field-level Masking (FM).

\paragraph{Pre-training Task for Payload}
Unlike the header, the payload lacks explicit continuity between its bytes.
This is primarily due to the nature of the payload, which often carries application-layer data or encrypted content.
During transmission, this data is typically fragmented, rearranged, or encrypted, causing the byte order to lose its direct connection to the original data structure.
Furthermore, encryption disrupts the continuity between bytes, increasing processing complexity.
Therefore, we directly use Masked Language Modeling based on Random Masking (MLM-RM).
Compared with field-level masking, random masking randomly selects $r$ tokens in the input token sequence.
The loss function is defined as:
\begin{align}
L_{\textit{MLM-RM}} = -\sum_{i = 1}^{r} \log \left(P\left(\textit{MASK}_{i} = { token}_{i} \mid {\theta}_{P} \right)\right)
\end{align}
where ${\theta}_{P}$ represents the set of parameters of the transformer encoder. RM means Random Masking.

\paragraph{Dynamic Masking for Model Pretraining}

Traditional pre-training models typically adopt static masking, where a fixed subset of tokens is selected for prediction during the data preprocessing stage \cite{etbert,PERT}.
These masked positions remain unchanged throughout model pre-training, leading the model to repeatedly encounter the same masking pattern.
This approach restricts the model’s exposure to diverse traffic information and encourages memorization of specific positions rather than learning generic traffic representations, ultimately limiting generalization.
To address this limitation, we adopt a dynamic masking strategy that generates masked positions dynamically during model pre-training.
This approach allows the model to observe varied byte combinations across training iterations, thereby improving the generalization and robustness of the learned traffic representations.

\subsection{Model Fine-tuning}
\label{sec:Model Fine-tuning}

The objective of model fine-tuning is to classify a given network flow into a specific traffic category.
Based on the obtained generic packet representation, we can perform different traffic classification tasks.
We utilize multiple linear layers and the softmax function to predict the probabilities of the network flow belonging to different traffic categories.
Note that there is no need to design separate models for each traffic classification task.
This is because traffic characteristics can be fully exploited through header-payload differentiated pre-training to generate a generic packet representation, which then supports various traffic classification tasks.

\section{Evaluation}

In this section, we first introduce the selected datasets, baselines, and the detailed implementation of TraGe. Subsequently, we evaluate TraGe on two traffic classification tasks. Our evaluation aims to answer four research questions:

\noindent  \textbf{RQ1.} Can TraGe outperform state-of-the-art (SoTA) traffic classification models?

\noindent  \textbf{RQ2.} What contribution does each TraGe component make?

\noindent  \textbf{RQ3.} How do parameter changes affect TraGe’s classification performance?

\noindent  \textbf{RQ4.} How does data sampling variability affect the performance stability of TraGe?

\subsection{Experiment Setup}

\paragraph{Datasets}
We select three datasets from diverse sources for pre-training and fine-tuning, including ISCX-VPN \cite{ISCX-VPN}, USTC-TFC \cite{USTC-TFC}, and CIC-IoT \cite{ciciot2022}.
The ISCX-VPN dataset captures traffic from various behaviors across multiple applications and services within VPN environments, such as browsing, email, audio, video, and file transfer.
The USTC-TFC dataset contains 10 categories of benign traffic and 10 categories of malicious traffic, primarily from smart grid infrastructures.
The CIC-IoT dataset is an IoT traffic dataset designed for behavioral analysis and intrusion detection.

We use a portion of these datasets for model pre-training and the remaining data for model fine-tuning.
Two traffic classification tasks on the ISCX-VPN dataset are performed, i.e.,  application classification and service identification.
The application classification and service identification tasks focus on encrypted traffic in VPN environments.
To evaluate the model's ability to classify applications and services in detail, 
we further categorize the ISCX-VPN dataset according to applications and services, i.e., ISCX-VPN (App) and ISCX-VPN (Service), ultimately forming a 17-class application classification task and a 12-class service identification task.
Table \ref{tab:sta} presents the specific traffic category distribution for the two traffic classification tasks.

\begin{table}[t!]
\renewcommand{\arraystretch}{1.1}
\caption{The Category Information of the Fine-tuning Tasks.}
\begin{center}
\resizebox{0.49 \textwidth}{!}{%
\begin{tabular}{clc}
\hline
Task &
Traffic Category &
\#label \\ \hline

\multirow{5}{*}{Application Classification} &
ftps, spotify, vimeo, hangout, &
\multirow{5}{*}{17} \\

&
ICQ, netflix, gmail, AIM, &
\\

&
YouTube, voipbuster, sftp &
\\ 

&
email-client, scp, skype, &
\\ 

&
Torrent, Facebook, Tor &
\\\hline

\multirow{4}{*}{Service Identification} &
Chat, VPN-Chat, VPN-P2P, &
\multirow{4}{*}{12} \\

&
P2P, Voip, VPN-Voip, FT, &
\\ 

&
VPN-FT, Email, VPN-Email, &
\\

&
Streaming, VPN-Streaming &
\\ 

\hline

\end{tabular}%
}
\label{tab:sta}
\end{center}
\vskip -2em
\end{table}

\paragraph{Implementation Details and Baselines}
We implement TraGe and all baselines by using Python 3.10.14 with libraries including NumPy 1.26.14, PyTorch 2.3.0, and CUDA 12.4.
For model pre-training, we use BERT \cite{BERT} as the transformer encoder.
We perform pre-training for 100,000 steps, with a learning rate of 1e-3.
During field-level masking, we apply a geometric distribution with parameters $p$ = 0.7.
Here, $p$ controls the shape of the distribution.
During model fine-tuning, we set the number of epochs to 10, with a learning rate of 2e-5.
In the stage of model fine-tuning, we select a maximum of 5 packets per flow for traffic classification. 
We randomly select at most 5000 flows from each traffic category for all tasks.
Each sampled dataset is then divided into the training set, the validation set, and the testing set according to the ratio of 8 : 1 : 1.

To ensure a fair comparison, we use three typical metrics, i.e., Precision, Recall, and F1-Score.
We select twelve baselines, including (1) four statistics features based models, i.e., FlowPrint \cite{FlowPrint}, AppScanner \cite{AppScanner}, XGBoost \cite{XGBoost}, and DTree \cite{DTree}; (2) four deep learning based models, i.e., FS-Net \cite{Fs-Net}, EBSNN-L \cite{EBSNN}, EBSNN-G \cite{EBSNN}, TFE-GNN \cite{TFEGNN}; and (3) four pre-trained models, i.e., PERT \cite{PERT}, ET-BERT \cite{etbert}, NetGPT \cite{NetGPT}, and YaTC \cite{YaTC}.

\subsection{Overall Classification Performance (RQ1)}
We show the classification performance of TraGe and all baselines on two traffic classification tasks in TABLE \ref{tab:overall_app} and TABLE \ref{tab:overall_ser}. 
Bold text indicates the best results, and underlined text indicates the second-best results among all models.
It can be seen that TraGe outperforms all baselines with significant margins across both tasks. 
As shown in Table \ref{tab:overall_app}, it obtains a Precision of 0.7517, a Recall of 0.7541, and an F1-Score of 0.7484 on the application classification task. Similarly, Table \ref{tab:overall_ser} shows that TraGe achieves a Precision of 0.9333, a Recall of 0.9335, and an F1-Score of 0.9331 on the service identification task.
These results surpass the second-best results (0.7421, 0.7187, 0.7182, 0.9250, 0.9192, and 0.9207) by 1.29\%, 4.93\%, 4.20\%, 0.90\%, 1.56\%, and 1.35\%, respectively.
Besides, we observe that the second-best results are all achieved by pre-trained models rather than statistics features-based and deep learning-based models.
Further, the results show that TraGe outperforms these four pre-trained models, with average performance improvements of 4.21\%, 4.08\%, 2.57\%, and 6.97\% on two tasks.
This is because TraGe fully exploits the characteristics of network traffic data to learn a generic packet representation, thereby supporting the best classification results on different traffic classification tasks.
In contrast, the baselines fail to fully account for these traffic characteristics, resulting in less generic representations that limit their classification performance.

\begin{table}[t!]
\renewcommand{\arraystretch}{1.1}
\centering
\caption{Classification Performance Comparison of Different Methods on Application Classification Task.}
\resizebox{0.4 \textwidth}{!}{%
\begin{tabular}{c|ccc}
\hline

Method &
Precision &
Recall &
F1-Score \\ \hline

FlowPrint \cite{DEGNN} &
0.5904 & 
0.4304 & 
0.4494 \\

AppScanner \cite{AppScanner} &
0.7289 &
0.5361 &
0.5803 \\

XGBoost \cite{XGBoost} &
0.5177 &
0.4151 &
0.4304 \\

DTree \cite{DTree} &
0.4844 &
0.4483 &
0.4462 \\ \hline

FS-Net \cite{Fs-Net} &
0.4990 &
0.3996 &
0.4060 \\

EBSNN-L \cite{EBSNN} &
0.7176 &
0.6691 &
0.6781 \\

EBSNN-G \cite{EBSNN} &
0.7285 &
0.6780 &
0.6882 \\

TFE-GNN \cite{TFEGNN} &
0.6720 &
0.6060 &
0.6180 \\ \hline

PERT \cite{PERT} &
0.7218 &
0.6971 &
0.6955 \\ 

ET-BERT \cite{etbert} &
0.7118 &
0.7027 &
0.6990 \\

NetGPT \cite{NetGPT} &
0.7336 &
\uline{0.7187} &
\uline{0.7182} \\

YaTC \cite{YaTC} &
\uline{0.7421} &
0.6746 &
0.6870 \\ \hline

\textbf{TraGe} &
\textbf{0.7517} &
\textbf{0.7541} &
\textbf{0.7484} \\

\hline

\end{tabular}%
}
\label{tab:overall_app}
\vskip -1.3em
\end{table}

\begin{table}[t!]
\renewcommand{\arraystretch}{1.1}
\centering
\caption{Classification Performance Comparison of Different Methods on Service Identification Task.}
\resizebox{0.4 \textwidth}{!}{%
\begin{tabular}{c|ccc}
\hline

Method &
Precision &
Recall &
F1-Score \\ \hline

FlowPrint \cite{DEGNN} &
0.7021 & 
0.6662 & 
0.6451  \\

AppScanner \cite{AppScanner} &
0.8599 &
0.7567 &
0.7913 \\

XGBoost \cite{XGBoost} &
0.7094 &
0.7160 &
0.7032  \\

DTree \cite{DTree} &
0.7026 &
0.7090 &
0.6997  \\ \hline

FS-Net \cite{Fs-Net} &
0.7161 &
0.6363 &
0.6418 \\

EBSNN-L \cite{EBSNN} &
0.9000 &
0.8672 &
0.8771 \\

EBSNN-G \cite{EBSNN} &
0.9013 &
0.8784 &
0.8845 \\

TFE-GNN \cite{TFEGNN} &
0.8597 &
0.8095 &
0.8214 \\ \hline

PERT \cite{PERT} &
0.9213 &
0.9148 &
0.9149 \\ 

ET-BERT \cite{etbert} &
0.9214 &
0.9182 &
0.9191 \\

NetGPT \cite{NetGPT} &
\uline{0.9250} &
\uline{0.9192} &
\uline{0.9207} \\

YaTC \cite{YaTC} &
0.9053 &
0.8565 &
0.8663  \\ \hline

\textbf{TraGe} &
\textbf{0.9333} &
\textbf{0.9335} &
\textbf{0.9331} \\

\hline

\end{tabular}%
}
\label{tab:overall_ser}
\vskip -1.3em
\end{table}

\subsection{Ablation Study (RQ2)}

In this subsection, we present the results of an ablation study of TraGe on the application classification and service identification tasks, as shown in TABLE \ref{tab:ablation}.
To streamline the presentation, we use the following abbreviations: ‘FM’ for field-level masking, ‘DM’ for dynamic masking.
When FM is not applied, random masking is used by default. Similarly, in the absence of DM, static masking is employed.

As shown in TABLE \ref{tab:ablation}, the results for both the application classification and the service identification tasks show that removing either field-level masking or dynamic masking significantly degrades model performance. Specifically, eliminating field-level masking leads to a 5.88\% decrease in F1-score for the application classification task, while removing dynamic masking causes a 1.62\% reduction. Similar trends are observed in both Precision and Recall. Overall, the results demonstrate that TraGe’s header-payload differentiated pre-training enables the formation of a generic packet representation, which in turn improves model classification performance across different traffic classification tasks.

\begin{table}[t!]
\renewcommand{\arraystretch}{1.1}
\centering
\caption{Ablation study of key components in TraGe on the Application Classification and Service Identification tasks.}
\resizebox{0.43 \textwidth}{!}{%
\begin{tabular}{l|ccc}
\hline

\multirow{2}{*}{Method} &
\multicolumn{3}{c}{{Application Classification}} \\

&
Precision &
Recall &
F1-Score \\ \hline

\textbf{Ours (TraGe)} &
\textbf{0.7517} &
\textbf{0.7541} &
\textbf{0.7484} \\ \hline

Ours w/o FM &
0.7318 &
0.7044 &
0.7044 \\

Ours w/o DM &
0.7489 &
0.7349 &
0.7363 \\

Ours w/o FM \& DM &
0.7234 &
0.7106 &
0.7086 \\ \hline \hline

\multirow{2}{*}{Method} &
\multicolumn{3}{c}{{Service Identification}} \\ 

&
Precision &
Recall &
F1-Score \\ \hline

\textbf{Ours (TraGe)} &
\textbf{0.9333} &
\textbf{0.9335} &
\textbf{0.9331} \\ \hline

Ours w/o FM &
0.9275 &
0.9226 &
0.9242 \\

Ours w/o DM &
0.9316 &
0.9302 &
0.9305 \\

Ours w/o FM \& DM &
0.9241 &
0.9240 &
0.9236 \\

\hline

\end{tabular}%
}
\label{tab:ablation}
\vskip -1.3em
\end{table}

\subsection{Parameter Sensitivity Analysis (RQ3)}

\begin{figure}[t]
\centerline{\includegraphics[width= 1 \linewidth]{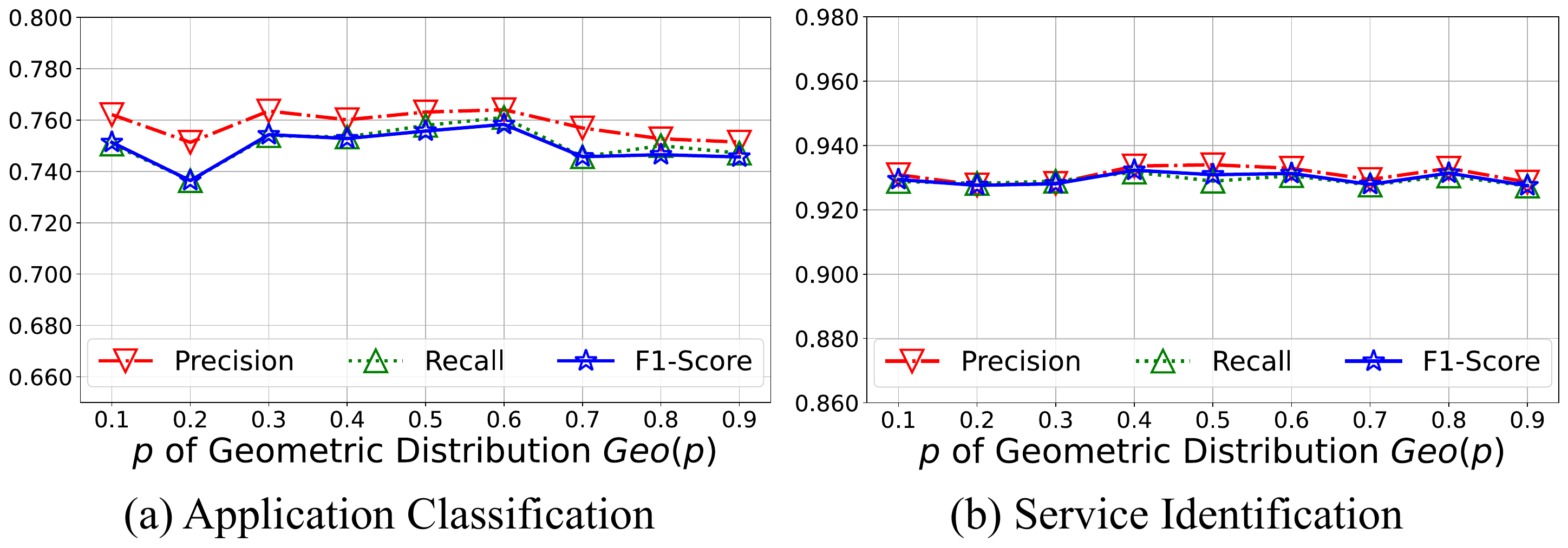}}
\vspace{-7pt}
\caption{Parameter analysis on field-level masking.}
\vspace{-5pt}
\label{fig:parameter}
\end{figure}

In this subsection, we evaluate the impact of various parameter settings in the header-payload differentiated pre-training on the effectiveness of TraGe.
In the pre-training of the transformer encoder, we utilize field-level masking to exploit the continuity of header fields.
Specifically, the length of the masked sequence is sampled from a geometric distribution $\ell \sim \operatorname{Geo} (p)$, where the parameter $p$ controls the masking length probability.
By adjusting the value of $p$, we aim to assess how sensitive the model is to the parameter of field-level masking.
As shown in Fig. \ref{fig:parameter}, TraGe exhibits stable classification performance on both application classification and service identification tasks across various values of $p$.
For instance, on the service identification task, the F1-Score fluctuates only slightly within the range of 0.9275 to 0.9323, yielding a maximum difference of just 0.0048.
These results demonstrate that TraGe is robust to changes in this masking parameter and that its effectiveness does not rely heavily on the configuration of the field-level masking.

\subsection{Sampling Variability Analysis (RQ4)}

\begin{figure}[t]
\centerline{\includegraphics[width= 1 \linewidth]{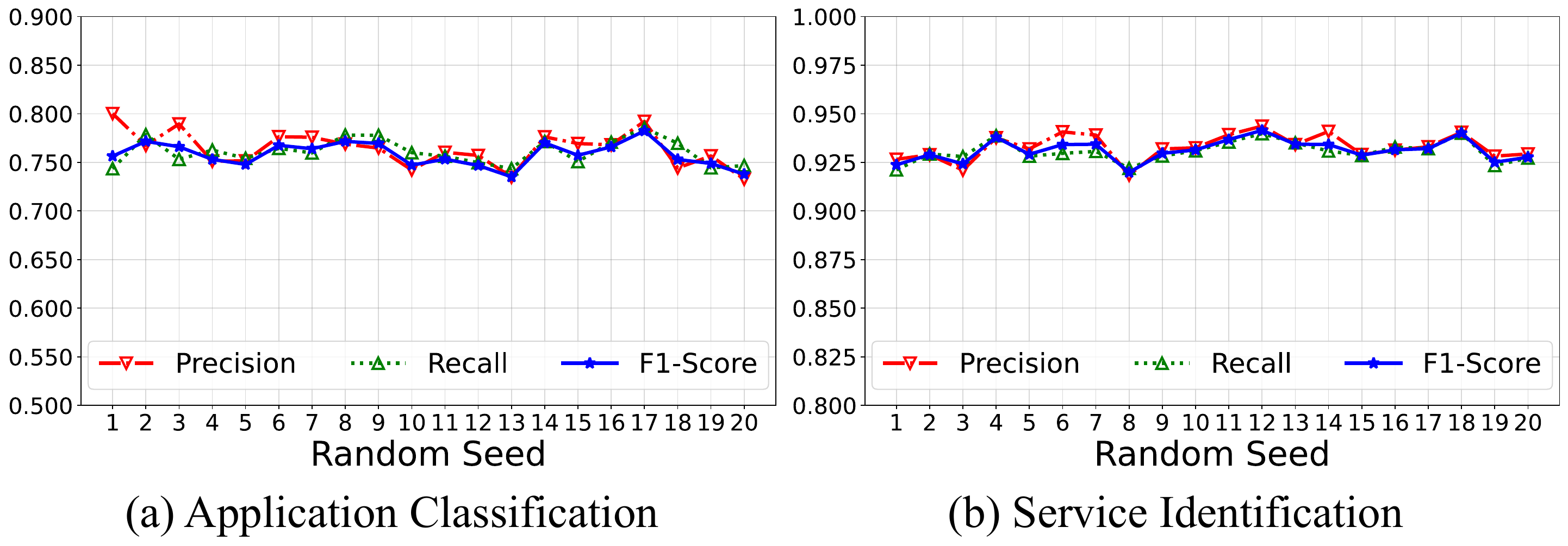}}
\vspace{-7pt}
\caption{Classification Performance under different sampling datasets.}
\vspace{-5pt}
\label{fig:random-seed}
\end{figure}

In this subsection, we evaluate the robustness of TraGe to variations in dataset sampling by evaluating its performance across multiple sampled datasets.
To simulate different training conditions, we randomly select up to 5,000 samples from each traffic category using a range of random seeds.
Specifically, we vary the random seed from 1 to 20, generating 20 distinct training and test set configurations.
Fig. \ref{fig:random-seed} presents the classification performance of TraGe under these different sampling datasets.
The results show that TraGe maintains stable performance across both the application classification and service identification tasks, regardless of the specific sampling instance.
For example, on the service identification task, the F1-Score varies only slightly between 0.9198 and 0.9414, with a maximum difference of just 0.0216.
This consistent performance across diverse sampled datasets demonstrates the robustness of TraGe to changes in sampling settings.

\section{Conclusion}

In this paper, we propose TraGe, a novel traffic classification model based on pre-training technology. 
Targeting the header-payload differences, TraGe effectively forms a generic packet representation adaptable to various traffic classification tasks.
An overall comparison, along with detailed evaluations—encompassing ablation study, parameter sensitivity analysis, and sampling variability analysis—is provided, demonstrating the effectiveness of TraGe.
The experimental results show that TraGe outperforms state-of-the-art methods on two traffic classification tasks.
Moreover, TraGe demonstrates strong robustness in both parameter sensitivity and sampling variability analyses.

\section*{Acknowledgment}

This work was supported in whole or in part, by National Natural Science Foundation of China (62372429, U24B6012 and U2333201), the Innovation Funding of ICT, CAS under Grant No. E461040, Pilot for Major Scientific Research Facility of Jiangsu Province of China (No.BM2021800).



\bibliographystyle{IEEEtran}
\bibliography{reference.bib}

\clearpage

\end{document}